\documentclass[12pt,a4paper]{article}      
\pdfoutput=1

\usepackage{bbm,amsthm,amsmath,amssymb,cancel,hyperref}
\usepackage{xcolor}
\usepackage[textwidth=6.3in,top=1.3in,bottom=1.3in]{geometry}

\newcommand{\acr}{\nonumber\\}
\newcommand{\ph}[1]{\phantom{#1}}
\newcommand{\mcl}[1]{\mathcal{#1}}
\newcommand{\itt}[1]{\intertext{#1}}

\newcommand{\mbf}[1]{\mathbf{#1}}
\newcommand{\abs}[1]{\left\lvert #1 \right\rvert}

\newcommand{\ess}[0]{\hspace{0.07em}}
\newcommand{\lss}[1]{\hspace{0.07em}#1}

\newcommand{\lamtil}{\widetilde\lambda}

\newcommand{\Gtil}{\widetilde G}
\newcommand{\fracs}[2]{{\textstyle\frac{#1}{#2}}}
\newcommand{\bx}{\mbf x}
\newcommand{\by}{\mbf y}

\newcommand{\bk}[0]{{\mathbf k}}

\newcommand{\dgznMt}[0]{P}

\newcommand{\dgzdCoeff}[0]{f}

\newcommand{\til}{\tilde}

\newcommand{\ra}{\rightarrow}

\def\baq{\begin{eqnarray}}
\def\eaq{\end{eqnarray}}

\newcommand{\beq}{\begin{equation}}
\newcommand{\eeq}{\end{equation}}

\newcommand{\al}{\alpha}

\newcommand{\si}{\sigma}

\newcommand{\lmk}{\left(}
\newcommand{\rmk}{\right)}
\newcommand{\lkk}{\left[}
\newcommand{\rkk}{\right]}

\newcommand{\newc}{\hat c}
\newcommand{\cC}{{\mathcal C}}

\begin{document}

\title{ \begin{center} \bf Inflationary dynamics of \\ kinetically-coupled gauge fields  \end{center}}

\vfill
\author{Ricardo Z. Ferreira\footnote{ferreira@cp3.dias.sdu.dk} ~ and~  Jonathan Ganc\footnote{ganc@cp3.dias.sdu.dk}
}

\date{}
\maketitle
\begin{center}
\vspace{-0.7cm}
{\it  CP$^3$-Origins, Center for Cosmology and Particle Physics Phenomenology}\\
{\it  University of Southern Denmark, Campusvej 55, 5230 Odense M, Denmark}

\end{center}
\vfill

\begin{abstract}

We investigate the inflationary dynamics of two kinetically-coupled massless $U(1)$ gauge fields with time-varying kinetic-term coefficients. Ensuring that the system does not have strongly coupled regimes shrinks the parameter space. Also, we further restrict ourselves to systems that can be quantized using the standard creation, annihilation operator algebra. This second constraint limits us to scenarios where the system can be diagonalized into the sum of two decoupled, massless, vector fields with a varying kinetic-term coefficient. 

Such a system might be interesting for magnetogenesis because of how the strong coupling problem generalizes. We explore this idea by assuming that one of the gauge fields is the Standard Model $U(1)$ field and that the other dark gauge field has no particles charged under its gauge group. We consider whether it would be possible to transfer a magnetic field from the dark sector, generated perhaps before the coupling was turned on, to the visible sector. We also investigate whether the simple existence of the mixing provides more opportunities to generate magnetic fields. We find that neither possibility works efficiently, consistent with the well-known difficulties in inflationary magnetogenesis. 

\end{abstract}

\tableofcontents

\section{Introduction}

The role of $U(1)$ gauge fields in Cosmology has been studied in many contexts. Undoubtedly, one of the most important goals has been the understanding of the magnetic fields that are pervasive in the observed universe. The magnetic fields in galaxies and clusters (e.g. \cite{Brandenburg:2004jv,Widrow:2011hs}), which appear today at the $\mu$G level, require an as-yet uncertain primordial seed. Furthermore, there are several claims of $\sim 10^{-15}$ G  magnetic fields in voids, possibly with coherence lengths of Mpc or larger\footnote{There are numerous other ways to constrain primordial magnetic field strength, notably but not exclusively from the CMB (see \cite{Durrer:2013pga} for more). One might also try to constrain other properties of a primordial magnetic field, e.g. its correlation with primordial density perturbations \cite{Ganc:2014wia}. At the moment, all measurements on primordial magnetic fields besides those from voids provide only upper bounds, although they must have been present at some level to seed the observed galactic fields.} \cite{Dolag:2010ni, Neronov:1900zz, Tavecchio:2010mk, Taylor:2011bn}. The proposed post-inflationary mechanisms, e.g. phase transitions or vorticities in the primordial plasma from second order perturbation theory (see \cite{Durrer:2013pga} for a recent review), are either too weak or small-scale to explain these observations.  Therefore, an inflationary scenario that explains these phenomena is quite desirable.

Since the conformality of the Maxwell action ordinarily prevents the generation of electromagnetic (EM) fields during inflation, most proposed inflationary magnetogenesis scenarios include terms that break conformality \cite{Turner:1987bw, Ratra:1991bn}. However, these proposals are now very constrained by a combination of considerations. Clearly, the EM field in such models must have a low enough energy to avoid backreaction on the inflationary dynamics \cite{Martin:2007ue,Kanno:2009ei,Subramanian:2009fu}. Also, one must ensure that the EM couplings remain perturbative during inflation, i.e. the ``strong coupling problem'' \cite{Demozzi:2009fu}\footnote{Shortly before we submitted this paper, preprint \cite{Tasinato:2014fia}, which has elements in common with \cite{Giovannini:2013rza}, was released, claiming to avoid the strong coupling problem by adding, in addition to the usual $f^2 F_{\mu \nu} F^{\mu \nu}$ term, the dimension 8 operator $i k_0 \partial^\mu\phi \partial_\nu\phi \left(\overline\psi\gamma^\nu\mcl D_\mu \psi\right)$, where $\phi$ is a fast-rolling scalar (not the inflaton) which has a large time derivative $\phi'$. In the scenario as outlined in the paper, the strong coupling problem seems to reappear in the 4-fermion interaction term $(e/f)^2\nabla^{-2}\left(\overline\psi\gamma^0\psi\right) \overline\psi\gamma^0\psi$, and also in the term $\left(e/ (\phi')^2 f\right)\,\partial_\mu\delta\phi\partial_\nu\delta\phi A^\mu \overline\psi\gamma_\nu\psi$ once backreaction constraints on $\phi$ are considered. One could consider adding additional operators to cancel out these problematic terms, though one would have to check that new problems do not arise. [We thank the author for discussions regarding this issue.]}. Furthermore, the energy in the EM fields generates a non-adiabatic pressure which causes the curvature perturbation to evolve outside the horizon; the perturbation generated is inherently non-Gaussian, so that excess EM field energy can generate unacceptably large non-Gaussianity \cite{Jain:2012ga,Bartolo:2012sd,Jain:2012vm,Shiraishi:2013vja,Fujita:2013qxa,Nurmi:2013gpa,Fujita:2014sna,Ferreira:2014hma}. Presently, there is no satisfactory mechanism which can explain the magnetic field observations and still satisfy all these constraints (unless inflation occurs at an extremely low energy scale \cite{Ferreira:2013sqa, Ferreira:2014hma}).

In this paper we consider the inflationary dynamics of two kinetically-coupled, massless gauge fields $A_a^\mu$ with field strengths $F_{ab}^{\mu\nu}$, $a,b\in\{1,2\}$, described by the action
\begin{align*}
  &S =  -\frac{1}{4} \int d^4x\,\sqrt{-g} \,\alpha_{ab}(\eta) F_a^{\mu\nu} F_{b\ess\mu\nu}\,.
\end{align*}
Such mixings were previously mentioned in a flat-space context in \cite{Holdom:1985ag} and in a magnetogenesis context, though quite briefly, in \cite{Barnaby:2012tk}. Suppose we associate $A_1^\mu$ with the Standard Model $U(1)$ field\footnote{Since this is before electroweak symmetry breaking, the field is, strictly speaking, the hypercharge $B_\mu$ but eventually will become part of the electromagnetic gauge field $A_\mu = \sin\theta_w A_\mu^3 + \cos\theta_w B_\mu$.} and $A_2^\mu$ with a field in the dark sector. One then wonders: if the dark field was excited before the mixing term $\alpha_{12}(\eta)$ was turned on, could these perturbations be transferred to the visible sector? Note that we can assume that there are no particles charged under the dark field, so it would not necessarily have a strong coupling problem, hence, dark magnetic fields would be rather easy to generate. Alternatively, we could ask: do the additional couplings in the action give us the ability to generate magnetic fields in the visible gauge field? Interestingly, if we ensure standard canonical commutation relations for the creation and annihilation operators associated with the gauge fields we are led to specific cases where the action can be decomposed into two decoupled gauge fields of the $f^2 F_{\mu \nu} F^{\mu \nu}$ type. As we will show this cases turn out to be inefficient for magnetogenesis.

The paper is outlined as follows. We begin in Sec. \ref{sec:action} by more explicitly outlining the scenario we explore. Next, we decouple the kinetic mixing term in the action by redefining the initial gauge fields as a linear combination of two other gauge fields. In Sec. \ref{sec:quantum-cons} we discuss issues related to the quantization of the fields. First, we investigate how a strong coupling problem in $A_1^\mu$ manifests itself in the transformed frame. Then, we try to quantize the fields and find that additional constraints appear if we want to write the fields in terms of the standard creation and annihilation operators. We proceed in Sec. \ref{sec:invest-quant-syst} by limiting ourselves to models that obey the aforementioned quantization conditions and focus on models with a constant diagonalization matrix, where the diagonalized action becomes that of two decoupled but non-conformal $U(1)$ gauge fields. We point out two specific classes of models where these conditions hold.  Then, in Sec. \ref{sec:cons-magn}, we investigate the possibility of generating magnetic fields in the visible sector by transferring magnetic fields from $A_2^\mu$ or by generating them directly in $A_1^\mu$ via the new couplings. As we conclude in the discussion in Sec. \ref{sec:disc}, we find in both cases that magnetic field generation is no more efficient that in the single-field case.

We have included an appendix \ref{sec:f2F2-case} where we review the main features of $f^2F^2$ models; in subsection \ref{sec:superh-lim-NO}, we write down the next-to-leading-order corrections to the long-wavelength limit of such models (which, as
far as we know, are not contained elsewhere). 

\section{Kinetic Mixing of Gauge Fields}
\label{sec:action}

A kinetic mixing between two massless gauge fields is described by the action \cite{Holdom:1985ag, Barnaby:2012tk} \footnote{Other types of couplings such as parity violating terms $ \lamtil(\eta) F_1^{\mu\nu} \Gtil_{\mu\nu}$, mass terms $ m^2_D(\eta) D^\mu A_{2\ess\mu}$ or  terms like $F_1^{\mu\nu} D^\mu A_{2\ess\mu}$ could also be considered although massive vectors are generically cosmologically unstable \cite{Himmetoglu:2008zp,Himmetoglu:2008hx,EspositoFarese:2009aj,Dimopoulos:2009vu}.}
\begin{align} \label{init-action}
  S =  -\frac{1}{4} \int d^4x\,\sqrt{-g} \, \alpha_{ab}(\eta) F_a^{\mu\nu} F_{b\ess\mu\nu}\,,
\end{align}
where $F_{a\ess\mu\nu}$ are the field strength tensors associated with two $U(1)$ gauge fields $A_{a\ess\mu}$, i.e. $F_{a\ess\mu\nu} \equiv \partial_\mu A_{a\ess\nu} - \partial_\nu A_{a\ess\mu}$, and $\alpha_{12} = \alpha_{21}$; we will generically use the indices $a, b$ to run over $\{1,2\}$ and to indicate to which gauge field we refer. We have broken the usual conformal symmetry of the Maxwell Lagrangian by allowing for couplings with other (unspecified) fields, which we parameterize by the functions $\alpha_{ab}(\eta)$ of the conformal time $\eta$.

We will work in a flat FRW universe, given by the metric
\begin{align*}
  ds^2 = a^2 (-d\eta^2 + d\mbf x^2)\,.
\end{align*}
It is convenient to perform a scale transformation $\overline{g}_{\mu\nu} = a^{-2} g_{\mu\nu} = \eta_{\mu\nu}$ to a flat metric which leaves the gauge fields $\{ A_{1\ess\mu}, A_{2\ess\mu}\}$ unchanged but where the action simplifies to
\begin{align} \label{eq:action}
  S =  -\frac{1}{4} \int d^4x\, \left[ \alpha_{11}(\eta) F_1^{\mu\nu} F_{1\ess\mu\nu} 
     + \alpha_{22}(\eta) F_2^{\mu\nu} F_{2\ess\mu\nu} + 2 \alpha_{12}(\eta) F_1^{\mu\nu} F_{2\ess\mu\nu} \right],
\end{align}
Electric fields $E^i$ and magnetic fields $B^i$ are defined in the standard way from the gauge fields as
\begin{align}\label{eq:dfn-E_B}
  &E^i_a = a^{-2} F_{a\lss{0i}}
   \underset{\text{gauge}}{\overset{\text{Coulomb}}{=}}
    -a^{-2} A_{a\ess i}'\,, &&\text{and} 
   && B^i_a = \fracs{1}{2}\epsilon_{ijl} a^{-2} F_{a\lss{jl}} 
     = i a^{-2} \epsilon_{ijl} k^j A_{a\ess l}\,,
\end{align}
where $E^i_1$, $B^i_1$ correspond to the Standard Model electric, magnetic fields, respectively, $\epsilon_{ijk}$ is the Levi-Civita symbol, and the second equation for the electric field is given in Coulomb gauge, where $A_{a\lss{0}}= \partial_i A_a^i = 0$, while the second one for the magnetic field is written in Fourier space. Heuristically, we see that electric fields come from the time derivative $A_{a\ess i}'$ of the gauge field while magnetic fields are stored in the field itself, i.e. $kA_{a\ess i}$. This understanding is useful when one wants to compare electric vs. magnetic field production because it translates simply into $A'_a$ vs. $kA_a$.

The equations of motion for the gauge fields, which follow from (\ref{eq:action}), are simply
\begin{align} \label{eq:EOM-A}
   &\partial_\mu \left(\alpha_{ab}(\eta) F_b^{\mu\alpha} \right) = 0 \,.
\end{align}

When analyzing magnetogenesis, we will consider $A_{1\ess\mu}$ as the Standard Model $U(1)$ field. However, the majority of our analysis applies quite generally. Until Secs. \ref{sec:quantum-cons} and \ref{sec:cons-magn}, our results are valid for any fields with action \eqref{init-action}. For those two sections, our results apply if the $A_{1\ess\mu}$ field has particles charged under its gauge group and if these interactions are to be understood perturbatively. One might be worried about observational constraints on dark field. However, it is phenomenologically straightforward to avoid cosmological signatures; in particular, one would merely need to ensure that its energy did not dominate the universe at any time and that, ultimately, its couplings to the Standard Model were sufficiently small or it acquired a large mass.

\subsection{Decoupling the kinetic mixing}
\label{sec:eliminating-kinetic-mixing}

We would like to decouple the kinetic terms in Eq. (\ref{eq:action}) through a transformation of the gauge fields, since this will make it easier to analyze the system. Note that \eqref{eq:action} can be written using matrices as
\begin{align} \label{eq:AcnWiMixMat}
  S = -\frac{1}{4} \begin{pmatrix} F_1^{\mu\nu} \  F_2^{\mu\nu} \end{pmatrix}
      \alpha \begin{pmatrix} F_{1\ess\mu\nu} \\ F_{2\ess\mu\nu} \end{pmatrix}\,,
\end{align}
where $\alpha$ is the coupling matrix given by
\begin{align} \label{eq:MixMat}
 \alpha \equiv \begin{pmatrix} \alpha_{11} &\alpha_{12} \\ \alpha_{12} & \alpha_{22} \end{pmatrix}\,.
\end{align}
A redefinition of the original gauge fields $A_{a\ess\mu}$ in terms of 2 new gauge fields $V_{a\ess\mu}$ with associated field strength tensors $X_{a\ess\mu\nu}$ can be done such that they are related to the original fields $A_{a\ess\mu}$ by
\begin{align} \label{eq:CanNorF}
  &\begin{pmatrix} A_{1\ess\mu} \\ A_{2\ess\mu} \end{pmatrix}
    \equiv P \begin{pmatrix} V_{1\ess\mu} \\ V_{2\ess\mu} \end{pmatrix}\,,
\end{align}
where $P$ is $2\times2$ mixing matrix. Thus, from \eqref{eq:AcnWiMixMat}, \eqref{eq:MixMat}, we see that diagonalizing the kinetic terms in the action is equivalent to choosing $P$ such that
\begin{align}\label{eq:dgzMixMat}
  P^T \alpha P = f^2\,,
\end{align}
where $f^2 \equiv \operatorname{diag}(\dgzdCoeff_1^2, \dgzdCoeff_2^2)$ is a diagonal matrix\footnote{Note that, sometimes, we also use the symbol ``$f$'' to refer to the coefficient of the kinetic term when speaking in general about $f^2F^2$ models. It should be clear from context which meaning is meant.}. Then the field strength tensors are given by,
\begin{align} \label{eq:CanNorFS}
  & \begin{pmatrix} F_{1\ess\mu\nu} \\ F_{2\ess\mu\nu} \end{pmatrix}
  = P \begin{pmatrix} X_{1\ess\mu\nu} \\ X_{2\ess\mu\nu} \end{pmatrix}
  + P' \left[\delta_{0\mu} \begin{pmatrix} V_{1\ess\nu} \\ V_{2\ess\nu} \end{pmatrix}
   - \delta_{0\nu} \begin{pmatrix} V_{1\ess\mu} \\ V_{2\ess\mu} \end{pmatrix}\right],
\end{align}
where ${}'\equiv \partial/\partial \eta$, which allows us to rewrite the Lagrangian as 
\begin{align*} 
  {\cal L}  
   &= -\frac{1}{4} \left[\dgzdCoeff_a^2 X_{a\ess\mu\nu} X_a^{\mu\nu}
    - 4 \begin{pmatrix} V_1^i \  V_2^i \end{pmatrix}\dgznMt^{\prime\,T} \alpha 
     \dgznMt \begin{pmatrix} X_{1\lss{0i}} \\ X_{2\lss{0i}} \end{pmatrix} 
   - 2 \begin{pmatrix} V_1^i \  V_2^i \end{pmatrix}\dgznMt^{\prime\,T} \alpha
     \dgznMt' \begin{pmatrix} V_{1i} \\ V_{2i} \end{pmatrix}\right] \,.
\end{align*}
As with $A_{a\lss0}$ in the original Lagrangian, $V_{a\lss{0}}$ do not have time derivatives and are thus not dynamical, reflecting the gauge freedom. 

Hereafter, we will work in Coulomb gauge, where we set $V_{a\lss{0}}= \partial_i V_a^i = 0$ and the Lagrangian becomes
\begin{align}\label{eq:ActionCoulomb}
  \mcl L
   &\begin{aligned}[t]
     =-\frac{1}{2} \Big[ & f_a^2 \big(\partial_\mu V_{ai}\big)^2
    - 2 \begin{pmatrix} V_1^i \  V_2^i \end{pmatrix}\dgznMt^{\prime\,T} \alpha
       \dgznMt \begin{pmatrix} V_{1i}' \\ V_{2i}' \end{pmatrix}
    - \begin{pmatrix} V_1^i \  V_2^i \end{pmatrix}\dgznMt^{\prime\,T} \alpha 
       \dgznMt' \begin{pmatrix} V_{1i} \\ V_{2i} \end{pmatrix}\Big]\,.
     \end{aligned}
\end{align}
Varying the Lagrangian with respect to the new gauge fields $\{V_{a\ess\mu}\}$ yields the following equations of motion
\begin{align}\label{eq:general-EOM}
     \begin{pmatrix}
       (\dgzdCoeff_1^2 V_{1i}')' - \dgzdCoeff_1^2 \partial^2 V_{1i} 
     \\ (\dgzdCoeff_2^2 V_{2i}')' - \dgzdCoeff_2^2 \partial^2 V_{2i}
     \end{pmatrix}
      + \begin{pmatrix} 1 \\ 1 \end{pmatrix} 
        \left[ \left( \dgznMt^{\prime\,T} \alpha \dgznMt - \dgznMt^{T} \alpha \dgznMt' \right)
           \begin{pmatrix} V_{1i}' \\ V_{2i}' \end{pmatrix} 
         + \dgznMt^T (\alpha \dgznMt')' \begin{pmatrix} V_{1i} \\ V_{2i} \end{pmatrix}
         \right] = 0\,,
\end{align}
which are, in general, a coupled system and complicated to solve.

\section{Quantum considerations}
\label{sec:quantum-cons}

In this section, we will discuss the validity of this theory within perturbative quantum field theory. First, we discuss how avoiding strongly coupled regimes due to changes in the effective charge of coupled particles constraints the parameter space. Second, we show that quantizing gauge fields with mixed time-derivative terms is non-trivial and that demanding that we can use the standard formalism further restricts the parameter space.

\subsection{Avoiding strongly coupled regimes}
\label{sec:strong-cplng}

Since we ultimately want to identify $A_{1\ess\mu}$ with the Standard Model $U(1)$ field, we suppose that there are particles charged under its gauge group with perturbative-strength couplings. On the other hand, there may be no particles charged under $A_{2\ess\mu}$.

Imagine a theory with a $U(1)$ gauge field $A_\mu$ and field strength tensor $F_{\mu\nu}$,
and a fermion $\psi$ charged under $A_\mu$. The Lagrangian which describes the coupling to fermions is\footnote{The particle $\psi$ need not be a spin $1/2$ fermion but the analysis is identical in other cases.}
\begin{align}\label{eq:cpld-action}
  \mcl L_\text{fermions} = -\frac{1}{4} f^2 F_{\mu\nu} F^{\mu\nu}
     + i \overline\psi (\cancel\partial - m)\psi +  g \overline\psi \gamma^\mu A_\mu \psi\,,
\end{align}
where $g$ is the charge of $\psi$ and $\gamma^\mu$ are the Dirac matrices. Initially, let us suppose $f=1$. If the charge $g$ is sufficiently large, perturbative quantum field theory would break down and standard quantum electrodynamics (QED) results no longer apply. Imagine instead that $g$ is of perturbative strength but $f$ is not unity; then, if we rewrite the action in terms of a ``canonically normalized''  field $\mcl A_\mu \equiv f A_\mu$, the action will look like \eqref{eq:cpld-action} with $f=1$ but $g \to g/f$. Again, perturbation theory breaks down if $f$ is sufficiently small such that $g/f$ becomes greater than order unity.

This same problem arises in \eqref{init-action} since a particle $\psi$ charged under $A_{1\ess\mu}$ becomes charged under $V_{a\ess\mu}$. To see this, simply note that the action contains the term
\begin{align}
  \mcl L \subset g \overline\psi \gamma^\mu A_{1\ess\mu} \psi 
   = g P_{11} \overline\psi \gamma^\mu V_{1\ess\mu} \psi 
     + g P_{12} \overline\psi \gamma^\mu V_{2\ess\mu} \psi\,,
\end{align}
where $P$ is the matrix defined in Sec. \ref{sec:eliminating-kinetic-mixing} that connects $A_{a\ess\mu}$ and $V_{a\ess\mu}$. Then, following the same logic as before, we see that $\psi$ has the effective charge $g P_{11}/\dgzdCoeff_1$, $g P_{12}/\dgzdCoeff_2$ under $V_{1\ess\mu}$, $V_{2\ess\mu}$, respectively\footnote{For the sake of this analysis, we have considered $\alpha$ as constant which can be seen as an instantaneous analysis of the system and its effective charges.}. Thus, since $g$ is of perturbative strength, this constraint translates into the condition
\begin{align}\label{eq:strong-cplng-constraint}
  &\frac{P_{1a}}{\dgzdCoeff_a}\lesssim 1\,. && \textit{(strong coupling constraint)}
\end{align}

The improvement of this setup compared to a single-field $f^2 F^2$ scenario is that both $P_{1a}$ and $f_a$ are time dependent functions and, therefore, even if $f_a$ is increasing during inflation, we are not necessarily in a strong coupling regime. Nevertheless, as we will see next, the standard quantization procedures will restrict us to scenarios no better than the single-field case.

\subsection{Quantization}
\label{sec:quantization}

We proceed by quantizing the vector fields using the standard methods. Demanding the consistency of this method will limit us to a subset of the coupling functions defined in the action \eqref{eq:action}.

We usually quantize vector fields $V_{a\ess i}$ by writing them as
\begin{align*}
  V_{a\ess i} (\eta,\bx)
   = \sum_{\sigma=1,2}\int \frac{d^3k}{(2\pi)^3} \epsilon_{\sigma,i}(\bk) 
     \left[a_{a\ess\sigma,\bk}  V_a (\bk, \eta) e^{i \bk\cdot\bx} 
       + a^\dagger_{a\ess\sigma,\bk}  V_a^*(\bk, \eta)e^{-i\bk\cdot\bx} \right],
\end{align*}
where $V_a$ are the mode functions of $V_{a\ess i}$, the basis vectors $e_{\sigma,i}$ satisfy
\begin{align*}
  &\epsilon_{\sigma,i}(\bk) \epsilon_{\sigma',i}(\bk) =\delta_{\sigma\sigma'}\,,
  && \epsilon_{\sigma,i}(\bk) k_i = 0\,, 
  && \sum_{\sigma=1,2} \epsilon_{\sigma,i}(\bk) \epsilon_{j,\sigma}(\bk) = \delta_{ij} - k_i k_j /k^2\, 
    \equiv \mathcal P_{ij}(\bk)\,,
\end{align*}
(the last condition ensures that we remain in Coulomb gauge), and the creation and annihilation operators satisfy the standard commutation relations
\begin{align}\label{eq:crtn_annih_oprts}
  \big[a_{a\ess\sigma,\bk}^{\ph{\dagger}},  a^\dagger_{a'\sigma',\bk'} \big]
    = (2 \pi)^3 \delta^{(3)}(\bk-\bk') \delta_{\si \si'} \delta_{aa'},
\end{align}
with all other commutators vanishing.

\newcommand{\vcfld}[0]{C}

To examine what models we can quantize in this manner, let us parameterize all possible interactions between two vector fields $\vcfld_a$ up to first-order time derivatives:
\begin{align}\label{eq:drv_qntzn-general-action}
  \mcl L &= \fracs{1}{2} M_{1'1'} (\vcfld_{1\ess i}')^2 
    + \fracs{1}{2} M_{2'2'} (\vcfld_{2\ess i}')^2 
    + M_{1'1} \vcfld_{1\ess i}' \vcfld_1^i + M_{2'2} \vcfld_{2\ess i}' \vcfld_2^i\acr
  &\ph{={}}+ M_{1'2'} \vcfld_{1\ess i}' \vcfld_2^{i\prime} 
   + M_{1'2} \vcfld_{1\ess i}' \vcfld_2^i + M_{12'} \vcfld_{1\ess i} \vcfld_2^{i\prime}
   + \mcl F(\vcfld_{1\ess i},\vcfld_{2\ess i})\,,
\end{align}
where $M_{xy}$ are time-dependent coefficients, and $\mcl F(\vcfld_a)$ is an arbitrary function of $\vcfld_a$ which is unimportant here. For such a Lagrangian the conjugate momenta associated with $C_1$, $C_2$ is, respectively, 
\begin{align*}
  \Pi^1_i &\equiv \frac{\delta L}{\delta \vcfld_{1\ess i}'} 
   = M_{1'1'} \vcfld_{1\ess i}' + M_{1'1} \vcfld_{1\ess i} + M_{1'2'} \vcfld_{2\ess i}' 
     + M_{1'2} \vcfld_{2\ess i}\,, \acr
  \Pi^2_i &\equiv \frac{\delta L}{\delta \vcfld_{2\ess i}'} 
    = M_{2'2'} \vcfld_{2\ess i}' + M_{2'2} \vcfld_{2\ess i} + M_{1'2'} \vcfld_{1\ess i}' 
       + M_{12'} \vcfld_{1\ess i}\,.
\end{align*}

Then we insist that our field algebra obeys the canonical commutation relations\footnote{Note that all commutators are taken at the same time and will be shown without coordinates, i.e. for the commutator of two operator $\mcl O_1$, $\mcl O_2$ we use
  $\left[\mcl O_1^i, \mcl O_{2\ess j}\right] 
    \overset{\text{means}}{\to} \left[\mcl O_1^i(\bx,\eta), \mcl O_{2\ess j}(\by,\eta)\right]$.}:
\begin{align}\label{eq:canon-cmmtn}
  &[\vcfld_a^i, \Pi^b_j] = i \delta_{ab} \delta^i_{\perp j}\,, \quad \delta^i_{\perp j}(\bx-\by) \equiv  \int \frac{d^3k}{(2\pi)^{3}}  e^{i\bk\cdot(\bx-\by)} \mcl P^i_{\;j}(\bk), 
\end{align}
with all other commutators vanishing. This requirement is satisfied if, assuming $M_{1'1'}, M_{2'2'}\neq0$, the 3 following conditions on the action \eqref{eq:drv_qntzn-general-action} hold at all times:
\begin{align}\label{eq:qntzn-conds}
  &W(\vcfld_a,\bk) = i M_{a'a'}^{-1}\,, &&\textit{(Condition I)}\acr
  &M_{1'2'}=0\,, && \textit{(Condition II)}\acr
  &M_{12'} = M_{1'2}\,, && \textit{(Condition III)}
\end{align}
where $W(\vcfld_a,\bk)$ is the Wronskian defined as
\begin{align}
  W(\vcfld_a,\bk) \equiv \vcfld_a(\bk) \vcfld_a^{\prime *}(\bk) - \vcfld_a^*(\bk)\vcfld_a'(\bk)\,.
\end{align}
Note that Condition II tells us that we must quantize in the diagonalized frame. In decoupled scenarios, like the $f^2 F^2$ case, the Wronskian condition (Condition I) is automatically satisfied by all the solutions of the equation of motion; by contrast, here it acts as an extra constraint on the system.

We find that only a subset of all possible couplings $M_{ab}$ can be quantized using standard techniques. 
The complications arise because of the time dependence of the mixing terms, which makes the system generically non-diagonalizable. Techniques like Dirac brackets (since there are no second-class constraints) or the BRST formalism (since the problem is unrelated to gauge symmetry) are not helpful. It may be possible to use more involved methods such as changing the creation and annihilation operator algebra but such considerations are beyond the scope of this paper.
Effectively, as we will see in the next section, we find that the use of standard techniques lead us to cases where the kinetic part of the Lagrangian can be diagonalized into two decoupled gauge fields.

\section{Solvable Scenarios}
\label{sec:invest-quant-syst}

Let us apply the results from the previous section to the Lagrangian obtained in Sec. \ref{sec:action}. Comparing \eqref{eq:ActionCoulomb} with \eqref{eq:drv_qntzn-general-action}, we see that Condition II became satisfied when we decoupled the kinetic mixing term \footnote{Also, as assumed in the above derivation, $M_{1'1'}\to \alpha_{11}\neq0$ and $M_{2'2'}\to \alpha_{22}\neq0$.}. On the other hand, Condition I implies
\begin{align}
  W(V_a) \equiv V_a(\bk) V_a^{'*}(\bk) - V_a^*(\bk) V_a'(\bk) = i f_a^{-2}\,,\acr
  \itt{or, equivalently, using Eq. \eqref{eq:general-EOM}}
  \label{eq:wron-cond}
  \left(P^T (\alpha P')'\right)_{12} V_1^* V_2 
    - \left(P^T (\alpha P')'\right)_{21} V_1 V_2^* = 0\,.
\end{align}
Finally, Condition III requires $P^{\prime T}\alpha P$ to be symmetric, i.e. 
\begin{equation} \label{eq:sym-cond}
P^{\prime T}\alpha P = P^T\alpha P'.
\end{equation}

Generically, an arbitrary matrix $P$ that satisfies \eqref{eq:dgzMixMat} can be written as
\begin{align*}
  \dgznMt = \til P\til f^{-1}U f\,,
\end{align*}
where $U$, $f$ are arbitrary orthogonal, diagonal matrices, respectively, and $\til P$ is the orthogonal matrix whose columns are the orthonormal eigenvectors of $\al$
\begin{align*}
 \til P^T \alpha \til P= \til f^2,
\end{align*}
and, therefore, $\til f^2 \equiv \operatorname{diag}(\til f_1^2, \til f_2^2)$ is the diagonal matrix with eigenvalues $\til f_1^2$, $\til f_2^2$ of $\alpha$. However, satisfying the quantization conditions (\ref{eq:wron-cond}, \ref{eq:sym-cond}) actually requires solving a coupled system of equations which also includes the equation of motion \eqref{eq:general-EOM}. By inspection one can easily see that matrices $P$ that satisfy
\begin{align}
  P'=0\,
\end{align}
are solutions to this system. Furthermore, it seems unlikely for other solutions to exist unless they are extremely contrived (though we have not explicitly proven this). Therefore, we are led to scenarios that can be diagonalized by a constant mixing matrix $P$. Note that, while $P'=0$ does not imply that the elements of $\alpha$ must be constant, only a subset of models with time-varying $\alpha$ can be diagonalized by a constant $P$. 

In this case, the Lagrangian simplifies considerably to the sum of two decoupled $f^2F^2$ scenarios 
\begin{align} \label{eq:SimpAcn}
  {\cal L}  
   &= -\frac{1}{4} \dgzdCoeff_a^2 X_{1\ess\mu\nu} X_a^{\mu\nu} 
    = -\frac{1}{4} \left(\dgzdCoeff_1^2 X_{1\ess\mu\nu} X_1^{\mu\nu} 
        + \dgzdCoeff_2^2 X_{2\ess\mu\nu} X_2^{\mu\nu}\right)\,.
\end{align}
Individually, these scenarios have been extensively studied in the literature \cite{Widrow:2011hs,Ratra:1991bn}, and we review some of their relevant properties in Appendix \ref{sec:f2F2-case}. Notably, the equations of motion in Fourier space are 
\begin{align}\label{eq:EOM-1fldCanon}
  &(\dgzdCoeff_aV^i_a)'' + \lmk k^2  
    - \frac{\dgzdCoeff_a''}{\dgzdCoeff_a} \rmk (\dgzdCoeff_a V_a^i) = 0
    \qquad\textit{(no sum over $a$)}\,.
\end{align}
Notice that we must also demand that both $\dgzdCoeff_1$, $\dgzdCoeff_2$ are positive in order to ensure healthy kinetic terms. We can straightforwardly find a set of solutions if we choose $P$ to be orthogonal, namely, $P=\til P$, $f=\til f$. Further imposing $\til P'=0$, we are left with two different cases:
\begin{itemize}
\item{\it Case 1: $\al_{11}(\eta)=\al_{22}(\eta)$}

In this case, the matrix $\til P$ has the form
\begin{align}\label{eq:P-al_eq_bet}
  \til P = \frac{1}{\sqrt2} \begin{pmatrix} 1 & 1 \\ 1 & -1 \end{pmatrix}\,,
\end{align}
while the eigenvalues are $\til f_1^2=\al_{11} + \alpha_{12}$ and $\til f_2^2= \al_{22} -\alpha_{12}$. This case includes the simplest scenario for transferring a field generated in $A_2$ to $A_1$, namely having $\alpha_{11}$, $\alpha_{22}$ constant but with a time-varying $\alpha_{12}$.

\item{\it Case 2: $\alpha_{12}(\eta) = c\big[\alpha_{11}(\eta)-\alpha_{22}(\eta)\big]$}

The second case is less simple. Here $\til P$ is given by
\begin{align}\label{eq:P-lam_prop_al_mi_bet}
   & \til P =
    \begin{pmatrix}
      \sqrt{\frac{2}{4+c^{-1}\kappa_+}}
      \begin{pmatrix} \fracs{1}{2} \kappa_+\\1 \end{pmatrix} & 
      \sqrt{\frac{2}{4-c^{-1}\kappa_-}}
      \begin{pmatrix} \fracs{1}{2} \kappa_-\\-1 \end{pmatrix}
    \end{pmatrix}\,,
\end{align}
where $ \kappa_{\pm}\equiv \sqrt{4+c^{-2}} \pm c^{-1}$, with $c$ an arbitrary constant, and the eigenvalues are now $\til f_1^2=  \alpha_{11} + \fracs{1}{2} \alpha_{12} \kappa_-$ and $\til f_2^2= \alpha_{11} - \fracs{1}{2} \alpha_{12} \kappa_+$.
\end{itemize}

\section[Consequences for magnetogenesis]{Consequences for magnetogenesis\footnote{Spacial indexes are omitted in this section.}}
\label{sec:cons-magn}
In this Section we investigate whether action (\ref{init-action}) can have interesting consequences for inflationary magnetogenesis. We analyze two different cases. First in Sec. (\ref{sec:transf-corr}), we explore if one can transfer a magnetic (electric) field in the dark sector into a magnetic field in the visible sector. Afterwards, in Sec. (\ref{sec:gnt-w-kinetic-mixing}), we study whether one can generate a magnetic field directly in $A_1$ using the additional freedom given by the dark field.

\subsection{Transferring a correlation from the dark to the visible sector}
\label{sec:transf-corr}

Since we assume that no particles are charged under the dark gauge field and, thus, that it has no strong coupling problem in itself, it is straightforward to generate a parametrically large magnetic (or electric) field, given that the difficulties arise due to strong coupling considerations \cite{Demozzi:2009fu}. In what follows we investigate whether such a field can then be transferred to the Standard Model $U(1)$ field. We suppose the fields are decoupled until some time $\eta_\text{m}$, i.e. $\alpha_{12}^m=0$ and, hence, $\til f_1^m= \til f_2^m=1$ where 'm' denotes quantities evaluated at $\eta=\eta_\text{m}$. We will try to transfer the correlation using the more straightforward Case 1 derived in Sec. \ref{sec:invest-quant-syst} where we suppose $\alpha_{11}=\alpha_{22}=1$ and let $\alpha_{12}(\eta)$ be a function of time, though using Case 2 works similarly. Note that the specific value of $\alpha_{11}=\alpha_{22}$ makes essentially no difference in the calculation and is chosen for convenience.

We will assume that the relevant cosmological modes are well outside the horizon at $\eta^\text{m}$. Thus, we can find the solution of \eqref{eq:EOM-1fldCanon} for $V_a$ to lowest order in $k$, giving us (see Appendix \ref{sec:f2F2-case})
\begin{align*}
    &V_a(\eta) \approx  V^\text{m}_a 
     + (\til f^\text{m}_{a})^2 V^{\prime\text{m}}_a \int_{\eta^\text{m}}^\eta \frac{d \eta'}{\til f^2_a(\eta')} ~.
\end{align*}
Then, using \eqref{eq:CanNorF} and \eqref{eq:P-al_eq_bet}, we have that
{
\newcommand{\pmsp}{1.3ex}
\begin{align*}
  \begin{pmatrix} A_1 - A^\text{m}_{1} \\[\pmsp] A_2 - A^\text{m}_{2} \end{pmatrix}
    &\approx \frac{1}{2} \begin{pmatrix} 1 & 1 \\[\pmsp] 1 & -1 \end{pmatrix}
        \begin{pmatrix} (A^{\prime\text{m}}_1 + A^{\prime\text{m}}_2) \int_{\eta_\text{m}}^\eta \til f_1^{-2} d\eta'
           \\[\pmsp] (A^{\prime\text{m}}_1 - A^{\prime\text{m}}_2) \int_{\eta_\text{m}}^\eta \til f_2^{-2} d\eta'
        \end{pmatrix}\,.
\end{align*}
}
To avoid strong coupling, we must require $|\alpha_{12}| \lesssim 1/2$ which implies that the eigenvalues $\til f_i$ lie between $1/2 \lesssim \til f_i^2 \lesssim 3/2$, and therefore they cannot vary significantly. Assuming that $A_1^\text{m}$ is negligible, since, otherwise, we would already have a magnetic field in $A_1$, we find
\begin{align*}
  \frac{\rho_B (A_1)}{\rho^\text{m}_E (A_1) + \rho^\text{m}_E(A_2)} 
   &= \frac{k^2 |A_1|^2}{|A^{\prime\text{m}}_1|^2 + |A^{\prime\text{m}}_2|^2} 
   < k^2 \underbrace{
      \frac{\left(\abs{A^{\prime\text{m}}_1} + \abs{A^{\prime\text{m}}_2}\right)^2}{{|A^{\prime\text{m}}_2|^2 + |A^{\prime\text{m}}_2|^2}}} _{<2}
     \Big(\underbrace{\max \int_{\eta_\text{m}}^\eta \til f_i^{-2}(\eta') d\eta'}_{<2 (\eta-\eta_\text{m})}\Big)^2 \nonumber \\
  & <  8 \big[k (\eta - \eta_\text{m})\big]^2 \ll 1\,.
\end{align*}

The result is similar to the single-field $f^2F^2$ model where magnetic fields can be generated whose ratio with the electric fields goes as $(k\eta)^2$, but it is also known that this is not enough to generate sizable magnetic fields, e.g. \cite{Martin:2007ue}. Thus, this idea is not an improvement over the simpler scenarios.

\subsection[Kinetic Mixing from early times]{Kinetic Mixing from early times\footnote{Note that the arguments of this section are valid for all models where $P'=0$, not just those where $P$ is orthogonal, as in the simple cases in Sec. \ref{sec:invest-quant-syst}.}}
\label{sec:gnt-w-kinetic-mixing}

Alternatively, we could think of generating magnetic fields directly in $A_1$ given that the kinetic coupling introduces additional freedoms. In this scenario, we choose $P$ with $P'=0$. From Eq. \eqref{eq:CanNorF}, $A_1$ is related to $V_a$ as
\begin{align}
  A_1 = P_{11} V_1 + P_{12} V_2\,,
\end{align}
where $V_1$ behaves as a gauge field in a $f^2F^2$ model with $f=f_1$. In particular, we know (see Appendix \ref{sec:f2F2-case}) that the comoving magnetic energy in $V_1$ is
\begin{align*}
  \rho_{B}(V_1) \simeq  k^2 f_1^2 V_1^2\,.
\end{align*}
Ultimately, we are interested in the magnetic field value in $A_1$ at the end of inflation, where we expect that $\alpha_{11} \ra 1$ so that standard electromagnetism is recovered\footnote{In principle, one could generalize the argument here to cases where $\alpha_{11}$ goes to unity only after reheating, when the large ion density would prevent the magnetic field $B_1$ of $A_1$ from changing even if $\alpha_{11}$ is changing in time. In principle, though, the specifics of reheating could alter $B_1$ so they would have to be specified or restricted.}. At that time, the comoving energy stored in the visible magnetic fields is
\begin{align*}
  \rho^f_{B}(A_1) \simeq  k^2 A_1^2 \simeq  k^2 P_{11}^2 V_1^2\,.
\end{align*}
Then, the ratio between these two quantities is
\begin{align}
  \frac{\rho^f_{B}(A_1)}{\rho_{B}(V_1)}
    = \left(\frac{P_{11}}{f_1}\right)^2\,,
\end{align}
which is exactly the strong coupling ratio derived in Eq. \eqref{eq:strong-cplng-constraint} and which we required to be smaller than one at all times in order to avoid strong coupling regimes.  Furthermore, to generate primarily magnetic fields (see Appendix \ref{sec:f2F2-case}), $f_1$ must increase significantly, hence, by the end of magnetogenesis, this ratio becomes extremely small. Thus, the magnetic energy in $A_1$ is much smaller than the one in $V_1$ which means that the energy is elsewhere and the process produces magnetic fields very inefficiently.

\section{Discussion}
\label{sec:disc}

In this work we analyzed the inflationary dynamics of  two kinetically-mixed $U(1)$ gauge fields $A_{a\ess\mu}$ with time-dependent kinetic-term coefficients described by action \eqref{init-action} which we ultimately study in the context of magnetogenesis. In general, this coupled system has complicated dynamics, as clear from the general equations of motion \eqref{eq:general-EOM}. However, some constraints have to be satisfied when quantizing the system. If the field has charged particles, we must avoid a strongly coupled regime in order to use perturbative quantum field theory. We derive the generalized strong coupling constraints for the two fields in \eqref{eq:strong-cplng-constraint}. Furthermore, we verified that if we demand that the fields can be quantized using the standard creation, annihilation operator algebra, we arrive at the conditions given in \eqref{eq:qntzn-conds}. 
These set of constraints, turn out to be quite restrictive and compel us to look at models with a constant mixing matrix $P$, which correspond to the cases where the Lagrangian decomposes into two decoupled $f^2F^2$ models. There are two families of solutions which can be so diagonalized: if $\alpha_{11}=\alpha_{22}$ or if $\alpha_{12} = c(\alpha_{11} - \alpha_{22})$. For any such models, the dynamics are straightforward to analyze.

To understand the implications for magnetogenesis, we identified one of the gauge fields with the Standard Model $U(1)$ field, which, of course, has charged particles. One could then hope that the extra freedom from the second gauge field, which may not have strong coupling problems in itself, could have interesting consequences for magnetogenesis. We first considered whether it was possible to transfer a strongly enhanced dark field $A_{2\ess\mu}$ to the visible sector's magnetic field on superhorizon scales. Then, we investigated whether the kinetic coupling between fields from early times could generate magnetic fields in the visible sector. In both cases we found the processes to be inefficient in the sense that most of the energy will be either in the electric field or in the dark field. We conclude that this setup is not an improvement over the simple $f^2 F^2$ model. 

Note that we were significantly restricted in our choice of kinetic-term coefficients by the requirements of standard quantization outlined in Sec. \ref{sec:quantization}, which led us to consider only systems that can be diagonalized into two independent $f^2F^2$ models (i.e. the action \eqref{eq:SimpAcn}\,). Such systems share most of the problems of single-field magnetogenesis (albeit in a slightly different form). However, there are no fundamental problems with the models we have discarded by our quantization demands; in a sense, they are merely inconvenient because they require a non-standard procedure for quantization. For example, it may be possible to simply alter the creation, annihilation algebra of fields while still satisfying the canonical commutation relations. While such considerations are beyond the scope of this paper, they might have interesting implications for magnetogenesis.

Thus, aside from having a very low scale of inflation, we still have no effective models of inflationary magnetogenesis, which continues to be in tension with the requirements of seeding the magnetic fields in galaxies and clusters and/or explaining void magnetic fields.\\

\noindent \textbf{Acknowledgments:} JG benefited from some discussions with Cliff Burgess regarding quantization issues. Tomohiro Fujita provided helpful comments on our draft. We thank Gianmassimo Tasinato for discussing some aspects of his recent proposal. We particularly acknowledge the interest, advice, and suggestions of Martin Sloth during the development of the paper. JG received partial support from the Munich Institute for Astro- and Particle Physics (MIAPP) of the DFG cluster of excellence during the workshop ``Origin and Structure of the Universe''. The CP3-Origins centre is partially funded by the Danish National Research Foundation, grant number DNRF90. We would like to thank the Lundbeck foundation for financial support.

\ \\

\appendix

\begin{samepage}
\noindent{\LARGE \textbf{Appendix}}

\section{The $f^2 F^2$ case}
\label{sec:f2F2-case}

\end{samepage}

The $f^2F^2$ model is described, after using the same scale transformation as in Sec. (\ref{sec:action}), by the action 
\begin{align*} 
  S =  -\frac{1}{4} \int d^4x f^2(\eta) F^{\mu\nu} F_{\mu\nu}\,,
\end{align*}
where $F_{\mu\nu}=\partial_\mu A_\nu - \partial_\nu A_\mu$ is the field strength associated with the $U(1)$ gauge field $A_\mu$. The corresponding energy-momentum tensor is
\begin{align*}
  T_{\mu\nu} = -\frac{2}{\sqrt{-g}} \frac{\delta\left[\sqrt{-g}\mcl L\right]}{\delta g^{\mu\nu}}
    = f^2\left( g^{\alpha\beta} F_{\mu\alpha} F_{\nu\beta} - \fracs{1}{4} g_{\mu\nu} F_{\alpha\beta} F^{\alpha\beta}\right)\,.
\end{align*}

Going forward, we will work in Coulomb gauge, where $A_0 = \partial_i A^i = 0$. The energy density is
\begin{align*}
  T_{00} = \frac{f^2}{2}\left( A_i' A_i'+ A_{i,j} A_{i,j}\right)
   = \frac{f^2a^4}{2}\big( \vec{E}^2 + \vec{B}^2\big)\,,
\end{align*}
where $E^i$, $B^i$ are defined in \eqref{eq:dfn-E_B}. The equation of motion for the canonically normalized gauge field ${\cal A}_i= fA_i$ is \cite{Martin:2007ue}
\begin{align}\label{eq:generic_eqn}
  {\cal A}_i'' + \left(k^2-\frac{f''}{f}\right){\cal A}_i = 0\,.
\end{align}
We can then quantize $A_\mu$ as we did for $V_{a\mu}$ in Sec. (\ref{sec:quantization}), which translates into the initial condition ${\cal A}\to 1/\sqrt{2k} \, e^{-i k \eta}$ when $-k \eta \to \infty$. In the superhorizon limit, when $\left\lvert f''/f \right\rvert\gg k^2$, we can use the solution of Eq. (\ref{eq:generic_eqn}) to lowest order in $k$
\begin{align*}
  A \equiv \frac{\mcl A}{f} \approx \cC_1 + \cC_2 \int \frac{d\eta}{f^2}\,;
\end{align*}
the next order in $k$, which is important, for example, if one needs to match different solutions at a given time, is derived in the following subsection.

 \ \

\subsubsection*{The power law case}

If we suppose that $f(\eta)\propto |\eta| ^\kappa$ \cite{Martin:2007ue}, then, we can find an exact solution for Eq. (\ref{eq:generic_eqn})
\begin{align}\label{eq:al_power_law-mode-fcn}
  A \equiv \frac{\mcl A}{f}
   &= (|k \eta|)^{-\kappa+1/2} \left(c_1 J_{\kappa-1/2}(|k \eta|) + c_2 J_{-\kappa+1/2}(|k \eta|)\right),
\end{align}
where $J_n(x)$ are Bessel functions of the first kind. In the superhorizon limit, this can be approximated by
\begin{align} \label{finalA}
  &A \approx   D_1 + D_2 |k \eta|^{1-2\kappa}\,,
  &A'(z) \approx \frac{2k}{\kappa+\frac{1}{2}} D_1 |k\eta|
     + \fracs{1}{2}k (\kappa-\fracs{1}{2}) D_2 |k\eta|^{-2\kappa}\,.
\end{align}
where $D_1$ and $D_2$ are found, using the initial condition for $A$ and Eq. \eqref{eq:al_power_law-mode-fcn}, to be
\begin{align*}
  D_1 &\equiv {\frac{\sqrt{\pi}}{2^{\kappa+1/2}}} \frac{e^{-i\pi\kappa/2}}{\Gamma(\kappa +\frac12)\cos(\pi\kappa)}\,,
   && D_2 \equiv {\frac{\sqrt{\pi}}{2^{3/2-\kappa}}} \frac{e^{i\pi(\kappa+1)/2}}{\Gamma(\frac32-\kappa)\cos(\pi\kappa)}\,.
\end{align*}

As described in Sec. (\ref{sec:action}), we can, naively, associate $kA$ to the magnetic field and $A'$ to the electric field. Noting that $|k\eta|\to0$ as inflation proceeds, we see from \eqref{finalA} that we need $\kappa<0$ to generate primarily of magnetic fields. However, this also corresponds to an increasing coupling function $f$. If $A_\mu$ is the Standard Model $U(1)$ field, the requirement that $f$ is increasing and $f=1$ at the end of inflation means that it must have started with $f\ll1$. As discussed in Sec. \ref{sec:strong-cplng}, since the Standard Model has fields charged under $U(1)$, when $f\ll1$ the system is at strongly coupled regimes, where calculations within perturbation theory are not valid anymore.

\subsection{Superhorizon limit to next order in $k$}
\label{sec:superh-lim-NO}

Let us compute the first order corrections in $k^2$ to the solution of \eqref{eq:generic_eqn} in the small $k$ limit. 
Suppose that 
\begin{align}\label{eq:Y-1st-order-k}
  {\cal A} \approx \cC_1 f \lkk 1+k^2 m(\eta) \rkk
   + \cC_2 f \lkk 1+k^2 n(\eta)\rkk \int \frac{d\eta}{f^2}\,,
\end{align}
where $m(\eta)$, $n(\eta)$ are unknown. If we plug this into the Eq. \eqref{eq:generic_eqn} and keep only terms up to first order in $k^2$, we find
\begin{align*}
  &{\cal A}'' + \left(k^2-\frac{f''}{f}\right){\cal A} \acr
  &\hspace*{2em}\approx k^2 \left\{ \cC_1 f \left(m'' + 2 \frac{f'}{f} m' + 1\right)
     + \cC_2 \left[2 \frac{n'}{f} + \left(2 f' n' + f(1+n'')\right) \int \frac{d\eta}{f^2} 
      \right]\right\} =0 \,.
\end{align*}
Imposing that the coefficient of $\cC_1$ to be zero gives us
\begin{align*}
  m &= \newc_1 \left(\int \frac{d\eta}{f^2}\right) - 
     \left(\int \frac{d\eta}{f(\eta)^2} \int_l^\eta f(\eta')^2 d{\eta'}\right)
   \simeq - \int \frac{d\eta}{f(\eta)^2} \int_l^{\eta} f(\eta')^2 d{\eta'}\,,
\end{align*}
where the second equality follows by removing the $\newc_1$ through a redefinition of $\cC_2$; note that the limits of both integrals in this equation are arbitrary since any changes can absorbed into redefinitions of $\cC_1$ and $\cC_2$.
In the same way, imposing the coefficient of $\cC_2$ to be zero leads to
\begin{align*}
  n &= \newc_2 \int \frac{d\eta}{g^2} - \int \frac{d\eta}{g(\eta)^2}\int_{l'}^\eta g(\eta')^2 d\eta'
   \simeq - \int \frac{d\eta}{g(\eta)^2}\int_{l'}^\eta g(\eta')^2 d\eta'\,,
\end{align*}
where in the second equality we have absorbed the $\newc_2$ term into a redefinition of $\cC_1$ and $\cC_2$; also, we have defined $g \equiv f \int d\eta\, f^{-2}$. As before, the limits of both integrals in this ast equation are arbitrary.

Thus, the generic solution of $A$ up to $k^2$ corrections is 
\begin{align}
  A &\equiv \frac{\cal A}{f} 
   \approx  \cC_1 \left(1 - k^2 \int \frac{d\eta}{f^2} \int_l^{\eta} f(\eta')^2 d{\eta'}\right)
    + \cC_2 \left(1 - k^2 \int \frac{d\eta}{g^2}\int_{l'}^\eta g(\eta')^2 d\eta'\right) 
     \int \frac{d\eta}{f^2}\,,
\end{align}
where all limits of integration can be set arbitrarily.

\providecommand{\href}[2]{#2}\begingroup\raggedright\endgroup

\end{document}